\documentclass[%
 reprint,
 superscriptaddress,
 amsmath,amssymb,
 aps,
]{revtex4-2}

\usepackage{graphicx}
\usepackage{dcolumn}
\usepackage{bm}

\usepackage{color}
\usepackage{booktabs}

\begin{document}

\preprint{APS/123-QED}

\title{Tidal Formation of dark matter deficit diffuse galaxy NGC1052-DF2 by SIDM}

\author{Zhao-Chen Zhang$^*$}
\affiliation{Key Laboratory of Particle Astrophysics, Institute of High Energy Physics,
Chinese Academy of Sciences, Beijing 100049, China}

\affiliation{School of Physical Sciences, University of Chinese Academy of Sciences, Beijing, China}

\author{Xiao-Jun Bi$^\dag$}
\affiliation{Key Laboratory of Particle Astrophysics, Institute of High Energy Physics,
Chinese Academy of Sciences, Beijing 100049, China}

\affiliation{School of Physical Sciences, University of Chinese Academy of Sciences, Beijing, China}

\author{Peng-Fei Yin$^\P$}
\affiliation{Key Laboratory of Particle Astrophysics, Institute of High Energy Physics,
Chinese Academy of Sciences, Beijing 100049, China}

\date{\today}

\begin{abstract}
Observations have revealed a significant dark matter deficit in the ultra-diffuse galaxy NGC1052-DF2 (DF2).
It is widely accepted that the formation of this unique galaxy can be attributed to the tidal stripping of its host galaxy, NGC1052.
In this study, we simulate the evolution of a satellite system containing globular clusters (GCs) within an accreting host halo in the framework of self-interacting dark matter (SIDM).
Our simulation results suggest that the heightened tidal stripping resulting from DM self-interactions can give rise to the transformation of a conventional dwarf galaxy into a dark matter deficit galaxy resembling DF2.
By comparing the simulation results with identical initial conditions in both the standard cold dark matter (CDM) and SIDM models, we find that the latter is more likely to replicate the properties of DF2.
Furthermore, we demonstrate that a DF2 analog can also be produced on an orbit with a greater pericenter distance by increasing the strength of DM self-interactions.
This suggests that the issue of extreme orbital parameters can be mitigated by implementing the SIDM model.
The distributions of the GC population derived in our SIDM simulation are consistent with the observed characteristics of DF2.
For comparison, we also explored the potential for achieving GC distributions in the context of CDM.
\end{abstract}

\maketitle


\section{\label{sec:introduction}Introduction\protect}

Dark matter (DM) has a significant impact on the formation of structures across various scales, from dwarf galaxies to galaxy clusters \cite{White:1977jf}.
Consequently, the observation of these systems offers a valuable avenue for investigating the properties of DM.
Ultra diffuse galaxies (UDGs) are a class of galaxies characterized by their remarkably low luminosity \cite{vanDokkum:2014cea}.
Given the minimal interference from baryonic matter, these galaxies provide an excellent opportunity for conducting such investigations. 

It is generally understood that the dynamics of UDGs is predominantly governed by DM \cite{vanDokkum:2016uwg}.
However, a remarkable UDG, NGC1052-DF2 (DF2), identified in \cite{vanDokkum:2018vup}, appears to exhibit a significant deficiency in DM.
The inferred stellar mass of DF2 is $M_{\star}=2\times10^{8}~\rm M_{\odot}$, while the estimated DM mass enclosed within $7.6~\rm kpc$ is also $\mathcal{O}(10^{8})~\rm M_{\odot}$ based on the kinetic data of globular clusters (GCs) in the galaxy \cite{vanDokkum:2018vup}.
This mass ratio is two orders of magnitude lower than the stellar-to-halo mass relation \cite{Moster:2012fv, Behroozi:2012iw, Behroozi:2019kql}.
Subsequent to the discovery of DF2, another dark matter deficient galaxy (DMDG), NGC1052-DF4 (DF4) has been identified \cite{van2019second}.
DF4 shares similar age, size, and mass distribution characteristics with DF2.
Recent observations and distance analysis have confirmed the status of DF2 and DF4 as DMDGs \cite{danieli2020tip, shen2021tip}.


The tidal effect has been proposed as a potential mechanism for the formation of DMDGs \cite{Penarrubia:2010jk}.
Since DF2 and DF4 are satellite galaxies of the NGC1052 group, the tidal field of the host halo could strip off DM in the outskirts of their halos.
Conversely, the stellar components bound in the center experience much lower mass loss, resulting in a reduced DM-to-stellar mass ratio.
In terms of observation, deep dragonfly imaging suggests that DF2 exhibits some tidal features in its morphology, indicating that DF2 may have undergone tidal evolution \cite{keim2022tidal}.
However, simulations show that DMDGs similar to DF2 and DF4 are unlikely to form within the standard cosmology framework \cite{Haslbauer:2019cpl}.
This implies that general tidal fields may not exert a sufficiently strong influence, and the formation of such DMDGs would require mechanisms to enhance tidal effects significantly.

Another anomaly of DF2 is the spatial distribution of its globular cluster (GC) population.
Given that the stellar component in DF2 has an age of approximately $8.9~\rm Gyr$ \cite{fensch2019ultra}, the GCs would likely have an effective radius significantly smaller than that indicated by the observations, due to dynamical friction.
Similarly, the tidal effect is also a possible explanation for this issue.
The tidal force acting on GCs counteracts the effect of dynamical friction, resulting in a diffuse distribution of GCs.


A DMDG similar to DF2 can be replicated in the N-body simulation featuring an initially cored DM halo as shown in \cite{Ogiya:2018jww}.
Ref. \cite{Ogiya:2021wyz} incorporates the accretion of the host halo into simulations, and explores the evolution of the stellar component and GC population.
Both studies examine the standard cold dark matter (CDM) scenario, wherein DM particles are collisionless and a density core may be formed through baryonic feedback \cite{Governato:2009bg, Pontzen:2011ty, DiCintio:2013qxa}.
However, a cored halo generated by feedback may be more likely to be destroyed by tidal force than to form a DMDG, due to its diffuse stellar distribution at the initial moment \cite{Yang:2020iya}.

Cored halos may arise as a natural consequence of self-interactions among DM particles.
The self-interacting dark matter (SIDM) model \cite{Spergel:1999mh} permits the exchange of momentum and energy within the DM halo, resulting in the thermalization of the inner halo and the formation of a cored density profile \cite{Burkert:2000di, Dave:2000ar, Vogelsberger:2012ku, Rocha:2012jg}.
In addition to explaining the flat central density of nearby dwarf galaxies, SIDM can address various other observational discrepancies with the $\Lambda$CDM paradigm, such as the diversity problem of rotation curves \cite{Kamada:2016euw, Ren:2018jpt, Kaplinghat:2019dhn} and the too-big-to-fail problem \cite{Vogelsberger:2012ku, Zavala:2012us, Elbert:2014bma}.
The results of simulations for SIDM conducted in \cite{Yang:2020iya} suggest that SIDM is more likely to produce the formation of DF2 and DF4.
However, the corresponding orbits of these two DMDGs may have a low probability of occurrence based on the results of cosmological simulations \cite{Jiang:2014zfa, vandenBosch:2017ynq}. 


In this study, we conduct N-body simulations in the framework of SIDM for DF2, aiming to provide a more comprehensive analysis.
Compared with \cite{Yang:2020iya}, we consider the accretion of the host halo and investigate the evolution of the GC population in DF2.
We determine whether the tidal force exerted by an accreting host halo can transform a typical dwarf galaxy ensconced in a SIDM halo into a DMDG resembling DF2, and examine the impact of DM self-interactions on the tidal evolution of stellar and DM components.
Specifically, we demonstrate that the extreme orbital parameters required to reproduce DF2 can be alleviated by increasing the cross-section of DM self-interactions.
We also show that the evolution of the GC population in our SIDM simulation is compatible with observations.
Furthermore, we discuss certain initial conditions that may achieve the final distribution of GCs in the scenario of CDM.

This paper is organized as follows.
In Section~\ref{sec:simulaton}, we provide a detailed description of the N-body simulations conducted in our study.
In Section ~\ref{sec:results}, we demonstrate the results of our benchmark simulation, which yields a DMDG exhibiting observational properties consistent with DF2.
The alleviation of extreme orbital parameters and the realization of GC distribution in the scenario of CDM are discussed in Section~\ref{sec:discussion}.
Finally, we offer a summary of our results in Section~\ref{sec:conclusion}.

\section{\label{sec:simulaton}Simulation}

In our endeavor to replicate DF2, we simulate the evolution of a typical dwarf galaxy in the tidal field of its host halo.
Considering that the indicated age of stars in DF2 is $8.9\pm1.5~\rm Gyr$ \cite{fensch2019ultra}, we adopt the initial redshift of the simulation to be $z_{\rm i}=1.5$, consistent with the value used in \cite{Ogiya:2021wyz}.
This initial condition corresponds to a look-back time of $-9.54~\rm Gyr$, and approximately $0.6~\rm Gyr$ later, the satellite galaxy falls into the host system.
In order to reduce computational costs, we represent the accreting host halo using a time-varying analytical background potential.
The N-body satellite system, comprising DM, stars, and GCs, is simulated using actual particles.
The results of the full simulation incorporating the host halo particles for standard CDM can be found in \cite{Katayama:2023fpc}.

We perform a benchmark simulation that produces a DMDG resembling DF2 in nearly all its observed properties with moderate DM self-interaction strength and orbital parameters.
For comparison, we also conduct an equivalent simulation in the standard CDM scenario.
Furthermore, we utilize different cross-sections of DM self-interaction to examine their influence on the orbit of the satellite.

\subsection{\label{sub:sidm} SIDM implementation and numerical settings}

To conduct SIDM simulations, we modify the public N-body simulation code  \texttt{GADGET-2} \cite{Springel:2005mi, Springel:2000yr}, utilizing the method described in \cite{Robertson:2016xjh}.
The results obtained from our modified code are in agreement with those derived from a semi-analytical method  \cite{Kaplinghat:2015aga} for the same halo, thereby verifying its accuracy.
The DM self-interaction considered in our simulation involves elastic scattering with an effective constant cross-section.
However, the realistic cross-section of SIDM may depend on the relative velocity between DM particles.
Specifically, in our implementation, the cross-section in a single halo is constant, but may vary in halos of different scales which have different average relative velocities between DM particles.
Ref. \cite{Kaplinghat:2015aga} shows that the cross-section of some observed dwarfs ranges from $0.3~\rm cm^{2}/g$ to $10.2~\rm cm^{2}/g$, thus we assume this as a reasonable interval for our simulations.
In the benchmark simulation, we have adopted a cross-section of $\sigma/m=5.0~\rm cm^{2}/g$, and the maximum value we used for alleviating the orbital parameters is $\sigma/m=13.0~\rm cm^{2}/g$. 

In our study, the initial conditions of DF2 including DM and stellar particles are generated by the code \texttt{Spheric} \cite{Garrison-Kimmel:2013yys}.
All these particles have a mass of $1.0\times10^{4}~\rm M_{\odot}$, and the corresponding softening length is taken to be $40~\rm pc$, consistent with the relation between resolution and softening length given by \cite{vandenBosch:2018tyt}.
This level of resolution is considered sufficiently high, effectively mitigating any artificial effects on the concerned substructures \cite{vandenBosch:2017ynq}.

\subsection{Host system}

In this study, the host system is characterized by the DM halo with an analytical potential.
We model the DM halo of NGC1052 using a Navarro-Frenk-White (NFW) profile \cite{Navarro:1996gj},
\begin{equation}\label{NFW}
    \rho(r)=\frac{\rho_{\rm s}}{\frac{r}{r_{\rm s}}(1+\frac{r}{r_{\rm s}})^{2}}, 
\end{equation} 
where $\rho_{\rm s}$ and $r_{\rm s}$ represent the scale density and radius, respectively.
An NFW profile can also be equivalently described by its viral mass and concentration parameter.
The viral mass $M_{200}$ is defined as the mass enclosed within a radius $r_{200}$, in which the average density is 200 times the critical density.
The concentration parameter is defined as $c=r_{200}/r_{\rm s}$.

NGC1052 has a stellar mass of $M_{\star}=10^{11}~\rm M_{\odot}$ \cite{forbes2017sluggs}.
According to the stellar-halo mass relation provided by \cite{Moster:2012fv}, the DM halo mass of NGC1052 is estimated to be approximately $M_{\rm 200}=1.1\times10^{13}~\rm M_{\odot}$.
To model the accretion process, we iteratively update the mass and concentration parameter at each time-step from $z_{\rm i}=1.5$ to $z_{\rm f}=0$, considering the DM halo's accretion history \cite{Correa:2015kia} and the redshift dependent mass-concentration relation \cite{Ludlow:2016ifl}.
Roughly, the $(M_{200},c)$ pair of the host halo evolves from $(3.6\times10^{12}~\rm M_{\odot},4.8)$ at $z=1.5$ to $(1.1\times10^{13}~\rm M_{\odot},6.8)$ at $z=0$.
Note that although the two canonical relations used above are established in the context of standard CDM, they are still applicable in the SIDM scenario, as DM self-interactions predominantly impact the distribution of the inner halo rather than the overall halo structure.

\subsection{\label{sub:sat} Satellite system}

The observed stellar mass of DF2 is estimated to be $2\times10^{8}~\rm M_{\odot}$ \cite{vanDokkum:2018vup}.
Because stars are gravitationally bound to the center of the DM halo and exhibit strong resistance to tidal forces, we expect that the decrease in stellar mass after tidal evolution will not exceed an order of magnitude.
Therefore, we take the initial stellar mass to be $M_{\star}=3.5\times10^{8}~\rm M_{\odot}$, slightly larger than the final observed value.
We set the initial DM halo mass to be $M_{\rm 200}=7\times10^{10}~\rm M_{\odot}$, which approximately follows the stellar-halo mass relation provided by \cite{Behroozi:2019kql}.

Assuming that the core formation occurs after the initial time of simulation, we model the initial DM halo of the satellite system with an NFW profile.
The initial concentration parameter is taken to be $5.5$ at $z_{\rm i}=1.5$, in accordance with the canonical mass-concentration relation in \cite{Dutton:2014xda}.
We assume the initial stellar distribution of the satellite system follows a Hernquist profile \cite{Hernquist:1990be} given by,  
\begin{equation}\label{Plummer}
    \rho(r)=\frac{\rho_{\rm h}}{\frac{r}{r_{\rm h}}(1+\frac{r}{r_{\rm h}})^{3}}, 
\end{equation} 
where $\rho_{\rm h}=1.6\times10^{8}~\rm M_{\odot}kpc^{-3}$ and $r_{\rm h}=0.7~\rm kpc$.
Under this profile, the effective radius of the satellite galaxy is $R_{\rm e}=1.25~\rm kpc$, which satisfies the relation between effective radius and stellar mass in \cite{carleton2019formation}. 

For the GC setting, according to the methodology outlined in \cite{Ogiya:2021wyz}, we determine a radius $r_{\rm GC}$ to the center of DF2 and take its value to be $r_{\rm GC}=1.8~\rm kpc$ in our benchmark simulation.
Subsequently, we select 10 star particles with positions closest to $r_{\rm GC}$ as the initial GCs, and adopt their velocities and positions as the initial GC conditions.
The mass of the GC particle is set to be $10^{6}~\rm M_{\odot}$, which is consistent with the average GC mass in DF2 \cite{van2018enigmatic}. 

\subsection{Orbit}

In the context of the discussion on orbits, DF2 is regarded as a point particle in the host halo.
Given our assumption of a spherical potential for the host halo, the orbit of DF2 can be determined by its orbital energy $E$ and orbital angular momentum $L$.
Specifically, we use two dimensionless parameters.
The energy parameter is defined as $x_{\rm c}\equiv r_{\rm c}(E)/r_{200}(z_{\rm i})$, and the circularity parameter is defined as $\eta\equiv L/L_{\rm c}(E)$.
Here, $r_{\rm c}$ and $L_{\rm c}$ represent the radius and angular momentum of the circular orbit with the same energy, respectively. 

The circularity parameter $\eta$, which ranges from 0 to 1, provides an indication of the orbit's radial nature, with smaller values indicating a more radial orbit.
And a smaller value of $x_{\rm c}$ represents a more bound orbit.
Cosmological N-body simulations show that the distribution of $x_{\rm c}$ and $\eta$ exhibits a peak around 1.2 and 0.6 \cite{Jiang:2014zfa, vandenBosch:2017ynq}, respectively.
In our benchmark simulation, we adopt $x_{\rm c}=0.8$ and $\eta=0.3$.
The pericenter distance of this orbit corresponds to 7.5 percentile of the distribution given by \cite{vandenBosch:2017ynq}.
We also conduct simulations with other orbital parameters to investigate whether a corresponding self-interacting cross-section can reproduce DF2.
The orbits used in this work are outlined in Table \ref{tab:ot}.

\begin{table}
    \centering
    \begin{tabular}{cccccc}
        \toprule
        $x_{\rm c}$ & $\eta$ & $r_{\rm peri} (\rm kpc)$ & $v_{\rm peri} (\rm km s^{-1})$ & $r_{\rm apo} (\rm kpc)$ & $v_{\rm apo} (\rm km s^{-1})$ \\
        \midrule
        $0.8$ & $0.2$ & $13.0$ & $656$ & $245$ & $34.9$\\
        $0.8$ & $0.3$ & $20.1$ & $639$ & $242$ & $53.1$\\
        $0.8$ & $0.4$ & $28.1$ & $609$ & $239$ & $71.6$\\
        $0.8$ & $0.5$ & $38.1$ & $561$ & $234$ & $91.5$\\
        \bottomrule
    \end{tabular}
    \caption{The comprehensive orbit parameters of DF2 used in our simulations.}
    \label{tab:ot}
\end{table}

\subsection{\label{sub:simplification} Simplification of host system}

The utilization of an analytical background potential to model the host system leads to the absence of two effects: dynamical friction and evaporation caused by DM self-interaction.
The timescale of orbit decay due to dynamical friction is significantly longer than the age of the universe, indicating that dynamical friction has a small impact on the orbit of the satellite system \cite{Ogiya:2018jww, Yang:2020iya}.
Among all the orbits used in this work, the minimum pericenter velocity is $560~\rm km/s$.
Ref. \cite{Kaplinghat:2015aga} performs a fitting of the average particle relative velocities and DM self-interaction cross-sections within some observed galaxies.
Considering this relation, the self-interactions between host and satellite DM particles have a cross-section of no more than $0.5~\rm cm^{2}/g$ when the satellite system passes the pericenter.
This interaction strength is weak, thus the evaporation effect can be neglected \cite{Silverman:2022bhs}.

In this investigation, we neglect the potential formation of a core in the host halo.
In theory, the inclusion of an isothermal core in the host halo would diminish the tidal effect.
To discuss the impact of the DM core, we perform the following calculation.
The host DM halo exhibits an average relative velocity among DM particles of $507~\rm km/s$, corresponding to a cross-section of $0.6~\rm cm^{2}/g$ according to the results of \cite{Kaplinghat:2015aga}. The characteristic radius $r_{\rm 1}$, where the average scattering time per particle within the age of halo is equal to unity, is calculated to be $37.6~\rm kpc$.
Ref. \cite{Jiang:2022aqw} provides a mass profile which can approximately describe the core: 
\begin{equation}\label{core}
    M(r)=M_{\rm NFW}(r)\text{tanh}(\frac{r}{r_{\rm c}}), 
\end{equation}
where $r_{\rm c}=0.45r_{1}=16.9~\rm kpc$ represents an effective core radius.
The pericenter distance in the benchmark simulation is $20.1~\rm kpc$, where the ratio between the enclosed mass in the core and the NFW case is 0.83.
We perform a simulation identical to the benchmark one, except for assuming the presence of a $16.9~\rm kpc$ core in the host halo throughout the simulation, which is an upper limit of core radius.
The results show that key properties of the satellite system, including the enclosed dynamical DM and stellar mass within various radii, the effective radius of stars, and the velocity dispersion of stars, differ from those in the benchmark simulation by no more than 10 percent.
Therefore, our approach to modeling the host DM halo is well justified.

\begin{figure}[tp!]
\includegraphics[width=\linewidth]{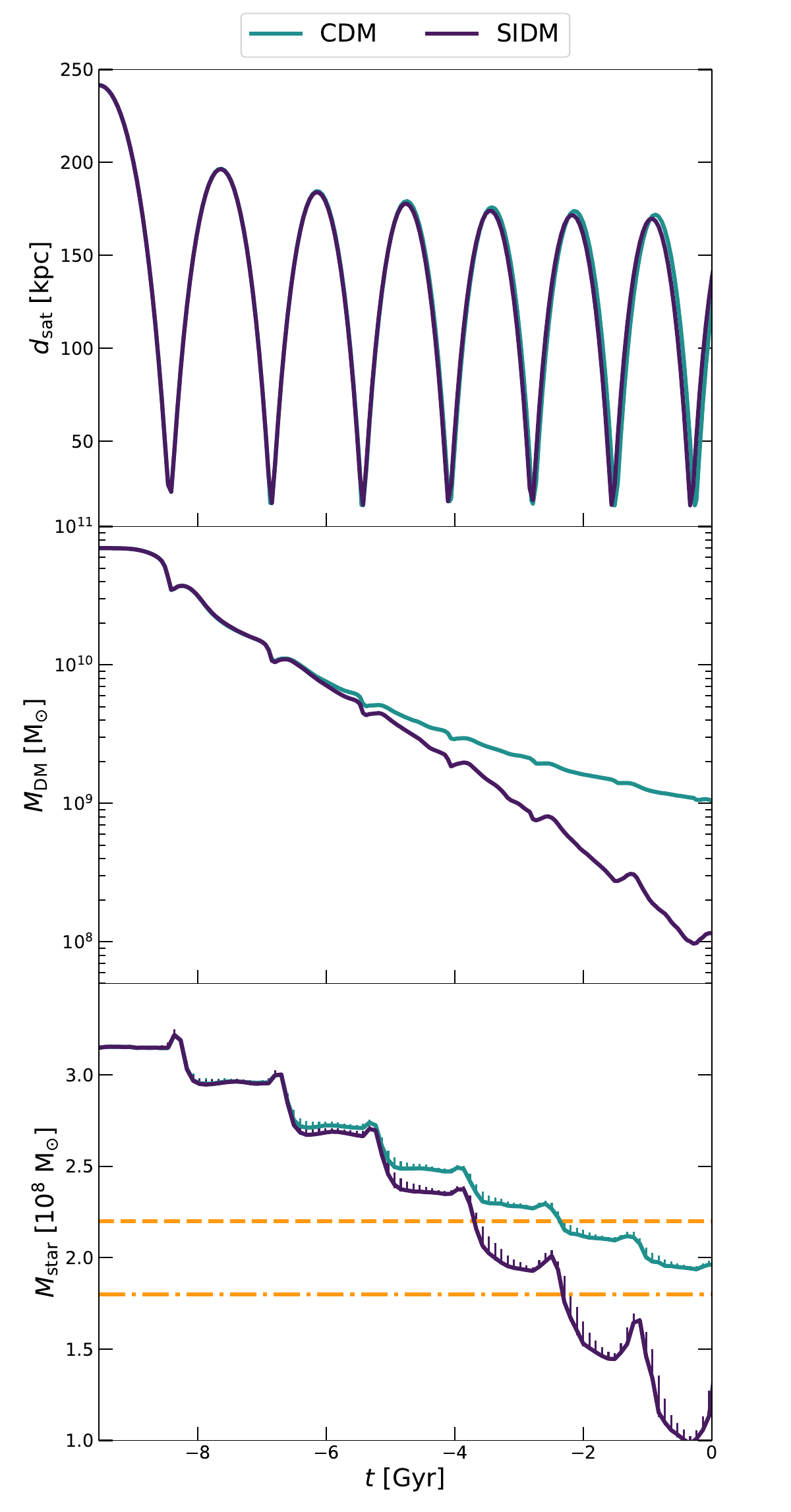}
\caption{\label{mass_loss}
The evolution of the position and mass loss of the satellite system.
The purple and green lines represent the SIDM and CDM results, respectively.
\textit{Top}: The distance from the center of the satellite system to the center of its host halo.
\textit{Middle}: The temporal evolution of the DM mass bound to the satellite system.
\textit{Bottom}: The detectable stellar mass over time.
The horizontal orange dash-dot line at $1.8\times10^{8}~\rm M_{\odot}$ represents the stellar mass inferred by the stellar population synthesis model, while the orange dash line at $2.2\times10^{8}~\rm M_{\odot}$ represents the stellar mass inferred by the mass-to-light ratio \cite{vanDokkum:2018vup}.
The error bars represent the 15 to 85 percentile range for the values obtained by projecting the stellar component using 100 random orientations.}
\end{figure}

\begin{figure*}[tp!]
\includegraphics[width=\linewidth]{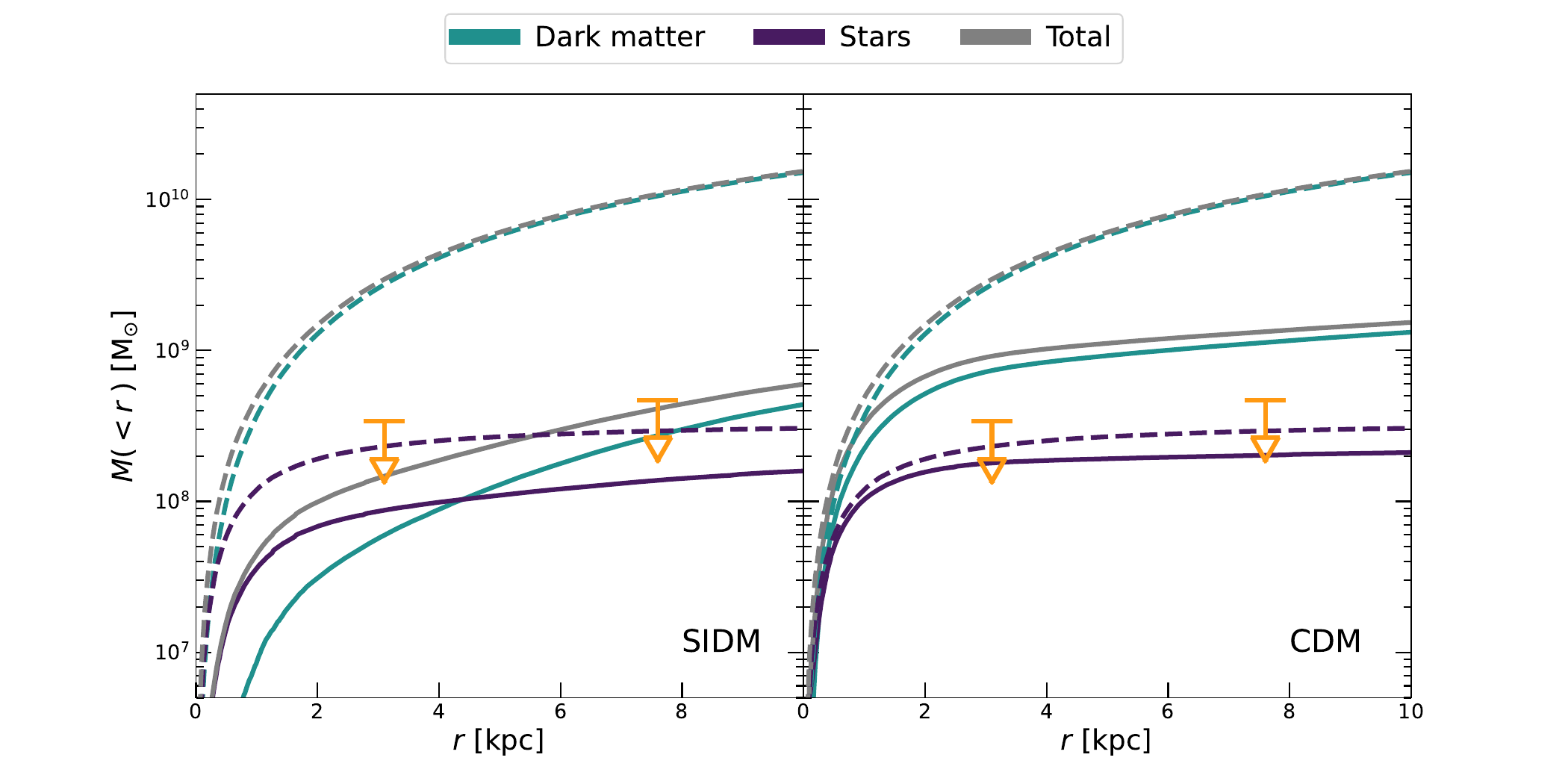}
\caption{\label{mass_profile}
The enclosed dynamical mass profile of the satellite system. The solid lines denote the mass profile at $t=-1.17~\rm Gyr$.
Note that the benchmark simulation results are most consistent with the observed data at this moment.
While the dashed lines denote the mass profile at the initial time for comparison.
The left and right panels represent the satellite halo in the SIDM and CDM scenarios, respectively.
The orange arrows represent the upper limits of the total dynamical mass enclosed within corresponding radii.}
\end{figure*}

\section{\label{sec:results}Results}

Using the initial conditions outlined in Section~\ref{sec:simulaton}, we conduct a benchmark SIDM simulation, the results of which are consistent with the observational data of DF2.
In this section, we analyze these simulation results and compare them with simulations in the standard CDM scenario under the same initial conditions.
Given that the infall time of DF2 can not be precisely determined, it is reasonable to regard any simulation snapshot within a small look-back time that satisfies all observational data as reconstructing a DMDG akin to DF2.
In the benchmark simulation, we identify that DF2 is replicated at $t=-1.17~\rm Gyr$.


\subsection{Mass loss}

The upper panel of Fig.\ref{mass_loss} shows the distance from the center of the satellite system to the center of its host halo.
In this analysis, we compute the center of mass of the 500 particles with the lowest potential energy to serve as the center of the satellite system.
The gradual decrease in apocenter distance with the orbital period is attributed to the accretion of the host halo.
Fig.\ref{mass_loss} indicates that there are essentially no differences in the orbital evolution of DF2 under the CDM and SIDM scenarios.
This can be attributed to the spherically symmetric nature of the satellite system and the isotropic DM scatter in the SIDM halo. 

The middle panel of Fig.\ref{mass_loss} illustrates the temporal evolution of bound DM halo mass.
Within the tidal field of the host halo, the DM component within the satellite system undergoes continuous stripping, resulting in a decrease in the DM mass.
To comprehend these findings, we can classify the SIDM halo into three layers based on radius: the inner region, middle region, and outer region.
Due to the low DM density, there is almost no DM self-interaction within the outer region.
The high-temperature DM particles in the middle region transfer kinetic energy to those in the inner region through self-interactions.
During the initial three pericentric passages, the main component that is stripped away is within the outer region, and due to the lack of particle scatter in this region, the mass change is essentially indistinguishable from that of a CDM halo.
Subsequently, as the DM particles in the inner region experience an increase in temperature caused by DM self-interactions, they move outward, resulting in a more shallow radial distribution of gravitational potential and a weakened resistance to tidal stripping.
This ultimately leads to more rapid mass loss compared to the CDM halo in the last four periods. 

The lower panel of Fig.\ref{mass_loss} depicts the evolution of detectable stellar mass over time.
Following the methodology outlined in \cite{Ogiya:2021wyz}, we project the 3-D spatial distribution of stellar particles along a specific direction of the line of sight, and calculate the mass of particles within $10~\rm kpc$ of the projected center as the detectable mass.
This is regarded as the mock observational mass in our simulations.
The central region of the satellite halo, where the stars reside, possesses a deep gravitational potential.
Consequently, the stars exhibit stronger resistance to tidal forces, resulting in a slower rate of mass loss compared to DM.
The SIDM halo has a more pronounced variation of DM gravitational potential due to the enhanced tidal stripping, thereby resulting in a faster loss of stellar mass compared to the CDM results.

Fig.\ref{mass_profile} shows the enclosed dynamical mass profile at $t=-1.17~\rm Gyr$ when the simulation results best match the observational data.
Using GCs as tracers, Ref. \cite{vanDokkum:2018vup} reports the upper limit of total dynamical mass within $3.1~\rm kpc$ to be $3.2\times10^{8}~\rm M_{\odot}$.
Considering the revision of velocity dispersion of GCs by \cite{van2018revised}, Ref. \cite{Yang:2020iya} argues that the upper limit within $7.6~\rm kpc$ should be $4.7\times10^{8}~\rm M_{\odot}$.
The total dynamical mass profile is distributed within the aforementioned upper limits in the SIDM scenario, while the mass distribution in the CDM halo exceeds these limits.
This discrepancy arises from DM self-interactions amplifying the tidal stripping, leading to a more substantial mass loss.

\begin{figure}[tp!]
\includegraphics[width=\linewidth]{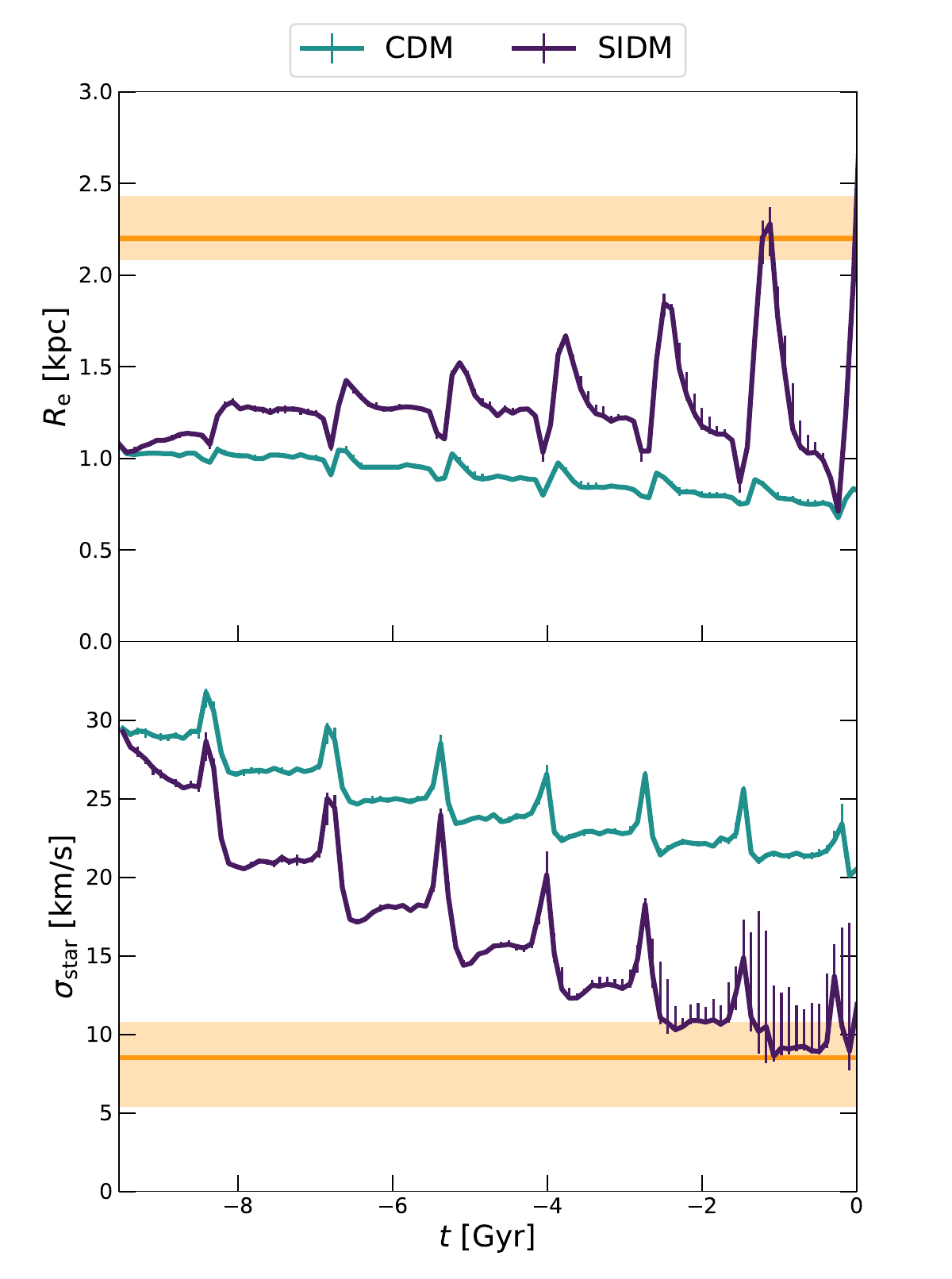}
\caption{\label{star}
The evolution of the stellar component in the tidal field.
The purple and green lines represent the SIDM and CDM results, respectively.
\textit{Top}: The effective radius $R_{\rm e}$ as a function of time.
The effective radius of stars in DF2 is inferred to be $R_{\rm e}=2.2~\rm kpc$ under the assumption that DF2 has a distance of $D=20~\rm Mpc$ \cite{vanDokkum:2018vup}, which is denoted by the horizontal orange line.
The lower and upper boundaries of the light orange region indicate $R_{\rm e}$ assuming the distance of DF2 to be $D=18.9~\rm Mpc$ \cite{cohen2018dragonfly} and $22.1~\rm Mpc$ \cite{shen2021tip}, respectively.
\textit{Bottom}: The line-of-sight velocity dispersion among the stars enclosed within the effective radius as a function of time.
The horizontal orange line and the light orange region represent the observed velocity dispersion of the stellar component in DF2, which is $\sigma_{\rm star}=8.5^{+2.3}_{-3.1}~\rm kms^{-1}$ \cite{Danieli:2019zyi}.
The error bars represent the 15 to 85 percentile range for the values obtained by projecting the stellar component using 100 random orientations.}
\end{figure}

\begin{figure}[tp!]
\includegraphics[width=\linewidth]{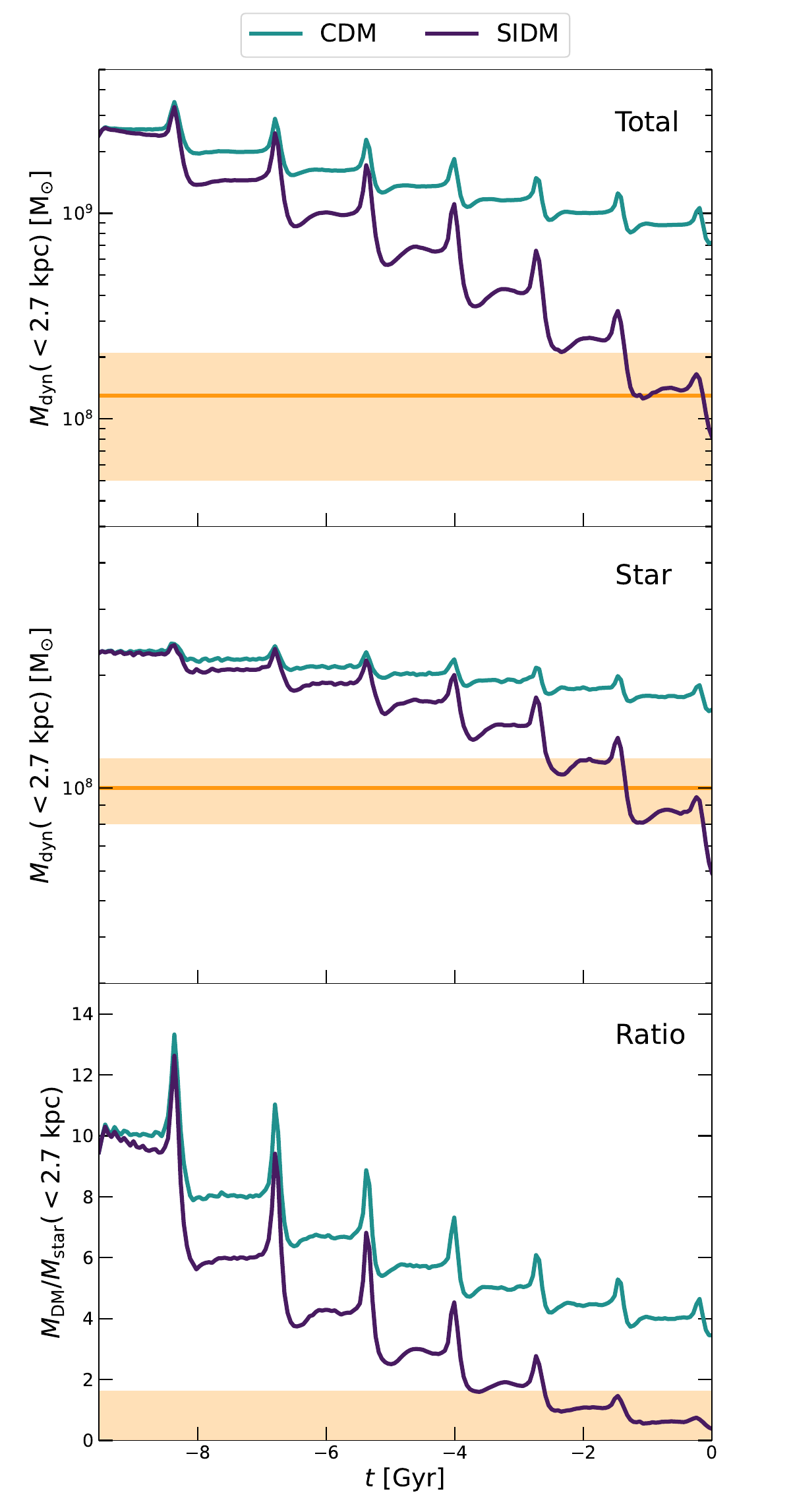}
\caption{\label{ratio}
The evolution of the dynamical mass enclosed within $R=2.7~\rm kpc$.
The purple and green lines represent the SIDM and CDM results, respectively.
\textit{Top}: The total dynamical mass enclosed within $R=2.7~\rm kpc$.
The orange line and light orange region represent the total dynamical mass inferred from observation, which is $(1.3\pm0.8)\times10^{8}~\rm M_{\odot}$ \cite{Danieli:2019zyi}.
\textit{Middle}: The dynamical mass of stars enclosed within $R=2.7~\rm kpc$.
The orange line and light orange region represent the inferred dynamical mass of stars, which is $(1.0\pm0.2)\times10^{8}~\rm M_{\odot}$ \cite{Danieli:2019zyi}.
\textit{Bottom}: The dynamical mass ratio of DM to stars enclosed within $R=2.7~\rm kpc$.
The light orange region represents the interval of the ratio deduced from the total and stellar mass mentioned above, ranging from 0 to 1.625.}
\end{figure}

\begin{figure}[tp!]
\includegraphics[width=\linewidth]{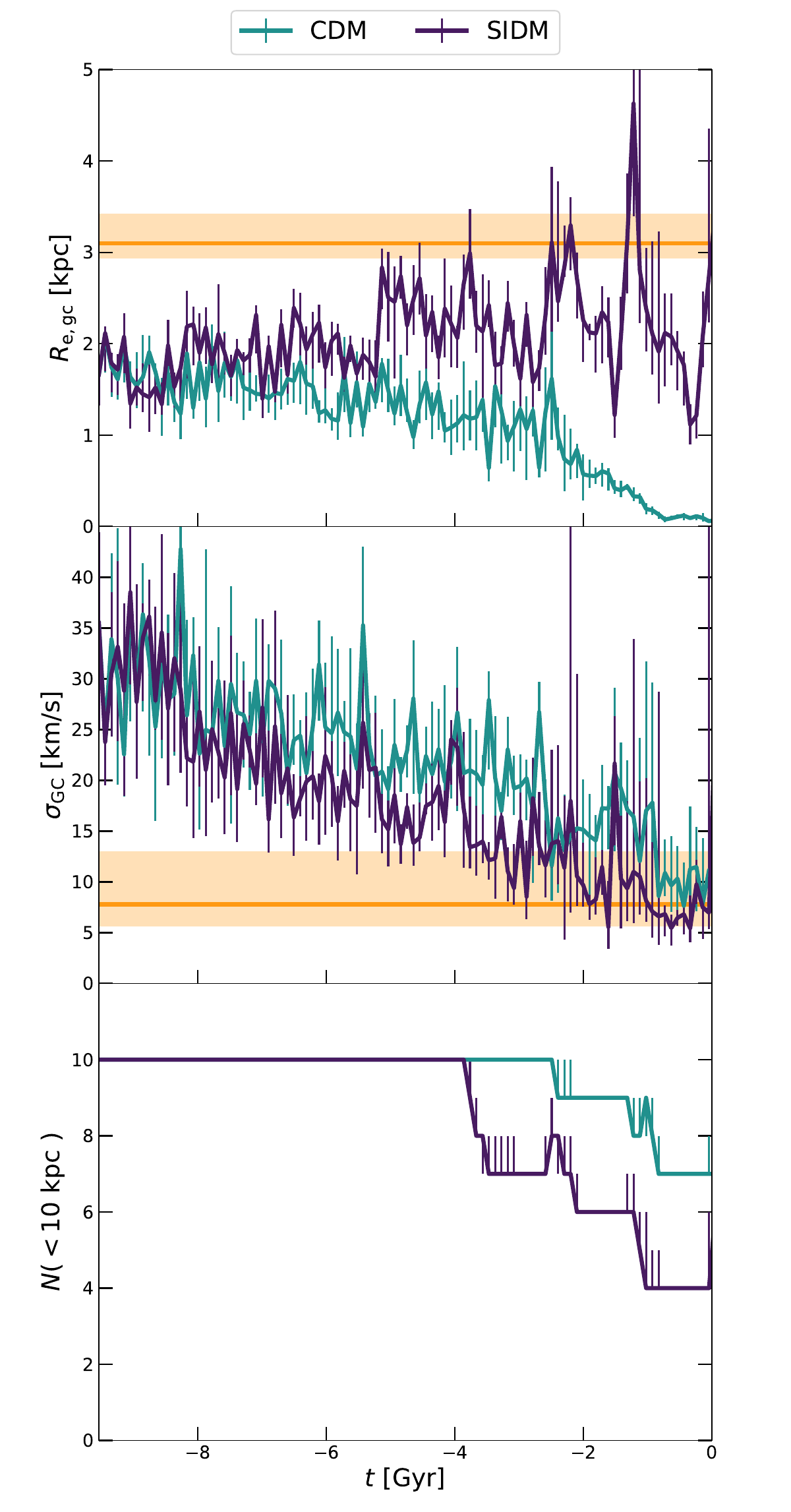}
\caption{\label{gc}
The evolution of GCs enclosed within $R=10~\rm kpc$ of the satellite system.
The purple and green lines represent the SIDM and CDM results, respectively.
\textit{Top}: The projected half-number radius of GCs within $R=10~\rm kpc$ as a function of time.
The horizontal orange line represents $R_{\rm e,gc}=3.1~\rm kpc$, as inferred from observations assuming the distance of DF2 to be $D=20~\rm Mpc$ \cite{vanDokkum:2018vup}.
The lower and upper boundaries of the light orange region indicate $R_{\rm e,gc}$ assuming the distance of DF2 to be $D=18.9~\rm Mpc$ \cite{cohen2018dragonfly} and $22.1~\rm Mpc$ \cite{shen2021tip}, respectively.
\textit{Middle}: The line-of-sight velocity dispersion among the GCs within $R=10~\rm kpc$.
The horizontal orange line and the light orange region represent the observed velocity dispersion of GCs, which is $\sigma_{\rm GC}=7.8^{+5.2}_{-2.2}~\rm kms^{-1}$ \cite{van2018revised}.
\textit{Bottom}: The number of GCs enclosed within $R=10~\rm kpc$.
The error bars represent the 15 to 85 percentile range for the values obtained by projecting the stellar component using 100 random orientations.}
\end{figure}

\subsection{Evolution of stellar component}

Fig.~\ref{star} illustrates the evolution of the effective radius $R_{\rm e}$ and line-of-sight velocity dispersion $\sigma_{\rm star}$.
The effective radius $R_{\rm e}$ is defined as the distance from the center of the satellite system in which half of the detectable stellar mass is enclosed, and $\sigma_{\rm star}$ is calculated from the star particles enclosed within $R_{\rm e}$.
The values of $R_{\rm e}$ obtained from projections in various directions show little variation, indicating that the star component enclosed within $R_{\rm e}$ roughly maintains a spherically symmetric spatial distribution.
This is because the central region of the galaxy is less affected by tidal forces and the DM self-scattering is isotropic.
Deep images of DF2 captured by different telescopes show that the isophotes near $R_{\rm e}$ have an ellipticity close to 0, verifying our inference \cite{keim2022tidal, golini2024ultra}.

The distribution of the stellar component is primarily influenced by two processes: tidal stripping and tidal heating.
During each pericenter encounter, the shrink of $R_{\rm e}$ and the increase of $\sigma_{\rm star}$ shown in Fig.~\ref{star} are both manifestations of tidal heating.
Subsequently, the stars expand and cool to a lower temperature owing to the negative heat capacity of a gravitational system.
Revirialization caused by tidal stripping also results in the cooling and expansion of the stellar component.
Overall, tidal effects lead to a more diffuse distribution and lower temperature of stars.
Note that the overall decreasing trend of $R_{\rm e}$ in the CDM simulation is caused by the pronounced mass loss of stars, and the star distribution still becomes more diffuse during the evolution.
Since the tidal stripping is enhanced by DM self-interaction, the SIDM halo exhibits a greater $R_{\rm e}$ and lower $\sigma_{\rm star}$, which are more consistent with the observation.
The distribution of stellar components once again indicates that SIDM is more likely to form a DMDG similar to DF2.

\subsection{Formation of DM deficiency }

The top and middle panels of Fig.~\ref{ratio} illustrate the total and stellar dynamical mass enclosed within $2.7~\rm kpc$ from the center of the satellite system, and the bottom panel shows the ratio of DM to star mass.
The $2.7~\rm kpc$ is the inferred 3D effective radius of DF2 \cite{Danieli:2019zyi}.
From the middle panel of Fig.~\ref{ratio}, we observe that the stellar mass enclosed within $2.7~\rm kpc$ eventually approaches half of the observed total stellar mass in the case of SIDM, once again demonstrating a good consistency between our simulation results and observational data.
The reduction of enclosed dynamical mass is the result of the increasing dispersion of both DM and stellar distribution.
The characteristic in which the total and stellar mass first increases and then decreases during each pericenter passage, is a manifestation of the combined effect of tidal heating and tidal stripping.

Because the tidal field primarily strips DM from the outer region, the revirialization makes more DM particles in the inner region move outward, leading to more rapid mass loss than the stellar component even in the central region.
Hence tidal stripping is able to reduce the DM-to-star mass ratio.
However, the tidal stripping in a CDM halo is not strong enough to generate a similar DM-to-star mass ratio as DF2.
In the case of SIDM, the DM particles in the inner region spontaneously diffuse outward due to heating by the ones in the middle region.
As a consequence, the SIDM halo exhibits a notably greater mass loss and decrease in DM-to-star mass ratio compared to the CDM halo in the central region.
In the final stage of evolution, although scatter between DM particles rarely occurs due to the low density in the SIDM halo, the shallower potential well attributed to reinforced tidal stripping still leads to a more pronounced DM mass loss.
Overall, in the SIDM scenario, the DM-to-star mass ratio decreases more rapidly and is more consistent with the inferred ratio interval.

\begin{figure*}[htp!]
\includegraphics[width=\linewidth]{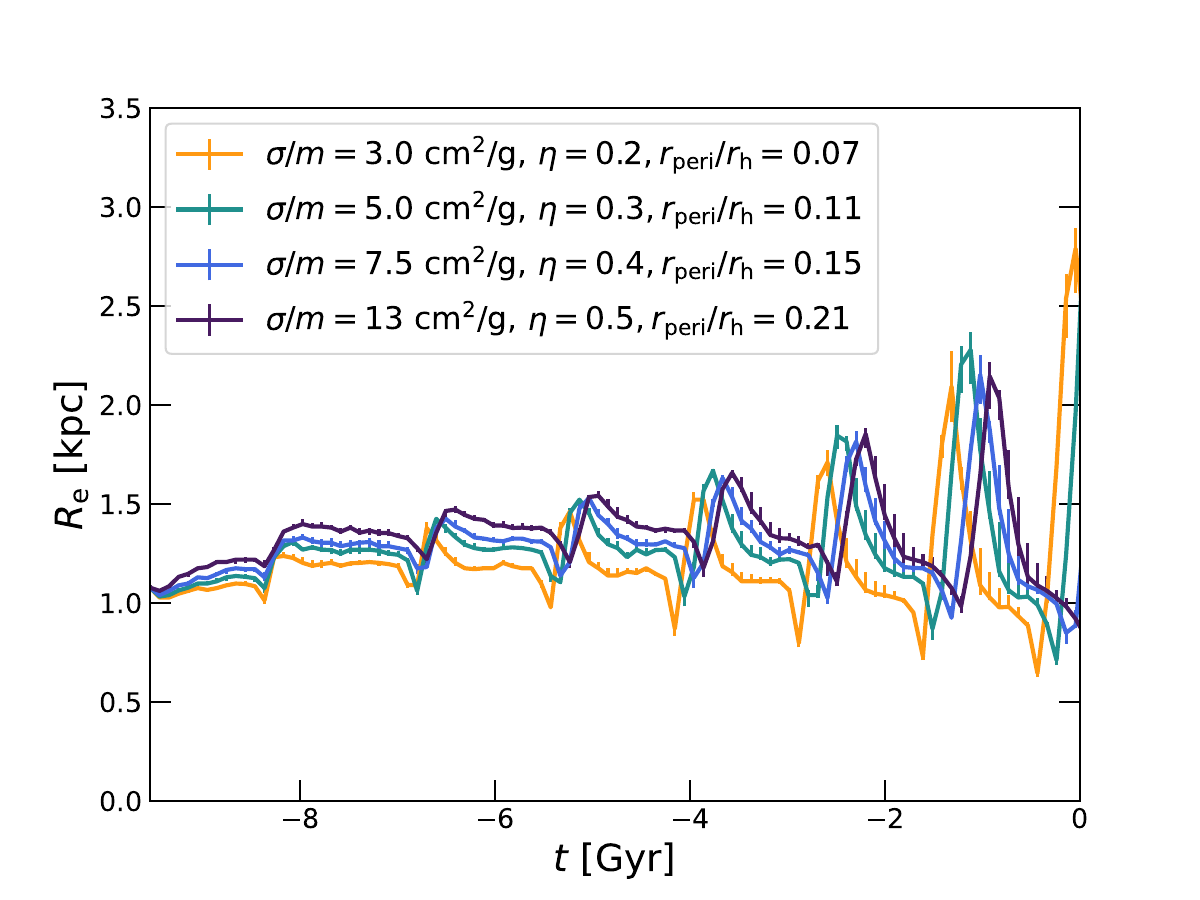}
\caption{\label{orbit}
The evolution of the effective radius in simulations with varying DM self-interacting cross-section $\sigma/m$ and orbital circularity $\eta$.
Lines of different colors represent simulations with different combinations of $\sigma/m$ and $\eta$, as indicated by the legend in the figure.
The green line with $\sigma/m = 5.0~\rm cm^{2}/g$ and $\eta = 0.3$ represents the benchmark simulation results analyzed in Section~\ref{sec:results}.
Note that the effective radius in all three simulations exhibits a similar peak value.}
\end{figure*}

\subsection{GCs}

Fig.~\ref{gc} shows the temporal evolution of the GC population.
The properties are all calculated using GC particles enclosed within $10~\rm kpc$ from the center of the satellite system, because these GCs are a reasonable proxy for the bound GCs \cite{Ogiya:2021wyz}.
Similar to stars, GCs also experience tidal heating and tidal stripping, which can lead to a dispersion in their spatial distribution and a decrease in temperature.
In addition, GCs are subject to dynamical friction, which makes them sink towards the center of the satellite system.
Because dynamical friction also leads to a reduction of the temperature, there is an overall decreasing trend in the velocity dispersion of GCs throughout the evolution.
However, in terms of spatial distribution, the influence of tidal effects and dynamical friction are opposite.
Due to varying initial velocity directions, certain GCs move toward the center, while others move outward.
The inner GCs are subjected to weaker tidal forces and tend to continue sinking inward under the influence of dynamical friction, while the outer ones experience stronger tidal forces and tend to continue expanding outward.
The spatial distribution of GCs is the combined result of the competition between tidal effects and dynamical friction.

In the case of SIDM, the influence of tidal effects is enhanced.
Consequently, under identical initial conditions, the spatial distribution of GCs in the SIDM halo is more diffuse than the one in the CDM halo.
Similarly, more GCs are stripped away in the scenario of SIDM.
When the simulation reproduces DF2 best ($t=-1.17~\rm Gyr$), there are $6\sim7$ GCs in the SIDM halo.
At a fixed time point, the number of GCs within a radius is assumed to be directly proportional to the initial number of GCs.
For the purpose of maintaining 10 GCs at $t=-1.17~\rm Gyr$, the initial number of GCs should be $14\sim17$.
Note the relation between the number of GCs and halo virial mass \cite{Burkert:2019ucv} suggests the satellite system includes 14 GCs initially, with which our prediction is a comparable value.
Therefore, our SIDM simulation is capable of generating a GC distribution similar to the one in DF2.

\section{\label{sec:discussion}Discussions}

\subsection{Mitigation of extreme orbital parameters} 

In addition to the benchmark simulation, we conduct simulations with varying orbital parameters, while keeping the energy parameters constant and modifying the circularity parameters.
Specific orbital parameters are adjusted as detailed in Table~\ref{tab:ot}.
As the circularity increases, the distance of the apocenter remains relatively constant, while the pericenter progressively moves further to the center of the host system.
Consequently, the tidal stripping of an orbit is weakened with an increase of circularity.
In order to achieve results similar to the benchmark simulation, we adjust the DM self-interaction strength to offset this influence.
Finally, we find three additional simulations which satisfy all observational data related to the distribution of DM and stars.
The evolution of the effective radius of these simulations is shown in Fig.~\ref{orbit} as an example.
Notably, the effective radius in all three simulations exhibits a similar peak value during each pericenter encounter.

\begin{figure}[htp!]
\includegraphics[width=\linewidth]{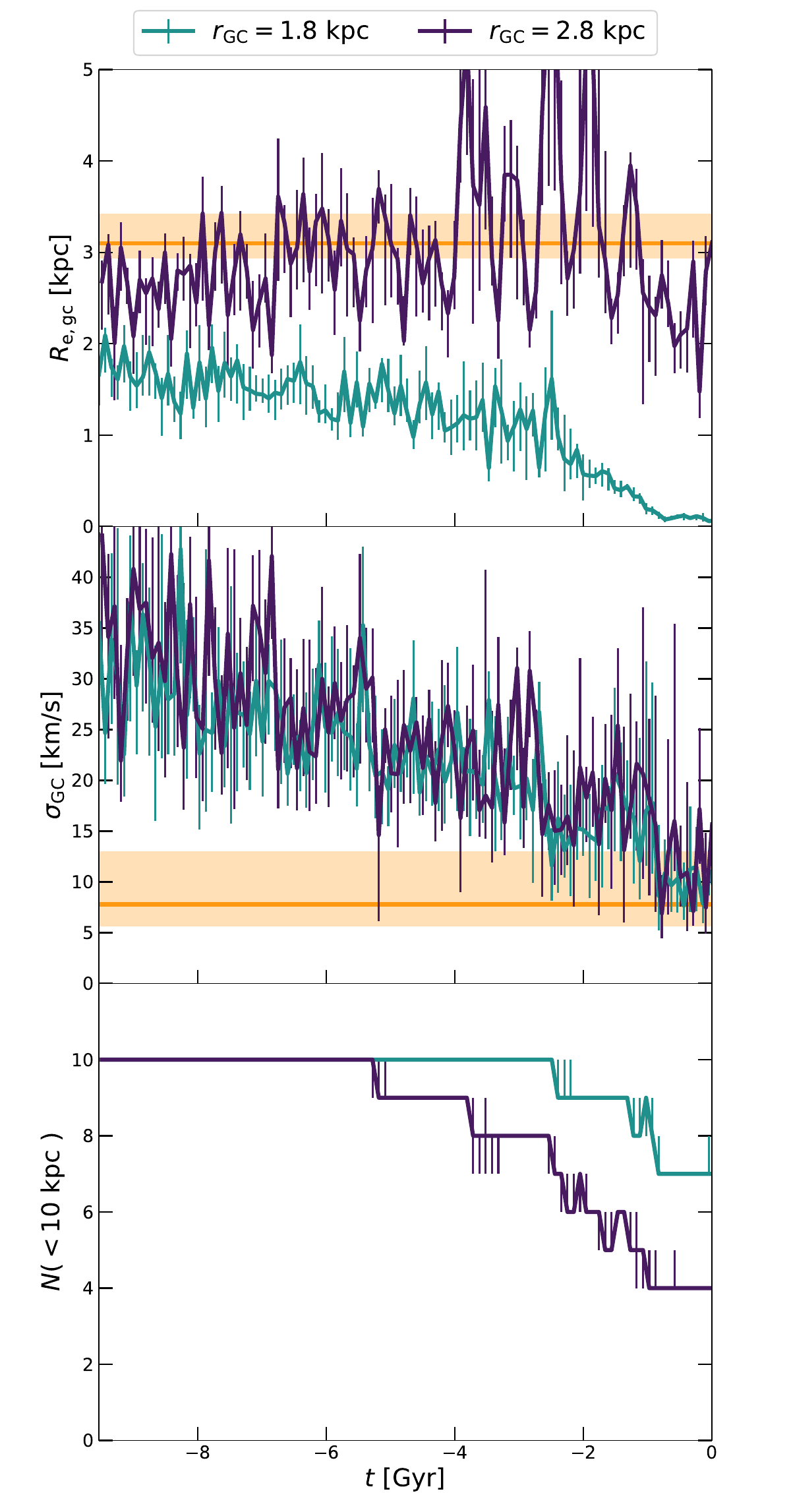}
\caption{\label{cdmgc} The temporal evolution of GCs enclosed within $10~\rm kpc$ in two CDM simulations with different $r_{\rm GC}$, represented by distinct colors.
This figure has the same setup as Fig. \ref{gc}.
\textit{Top}: The projected half-number radius of GCs within $10~\rm kpc$.
The light orange region represents the interval of $R_{\rm e,gc}$ inferred from observation, as discussed in detail in Fig.\ref{gc}.
\textit{Bottom}: The number of GCs within $10~\rm kpc$.
In the simulation with a larger $r_{\rm GC}$, more GCs are stripped away as the tidal evolution progresses.
The error bars represent the 15 to 85 percentile range for the values obtained by projecting the GCs using 100 random orientations.}
\end{figure}

In these simulations, the values of the cross-section $\sigma/m$ fall within a reasonable range, as discussed in Section~\ref{sub:sidm}, except for the simulation involving an orbit with $\eta=0.5$, where the value slightly exceeds the upper limit.
If the self-interaction is strong and the evolution time is sufficiently long, the temperature of the inner region of the DM halo becomes higher than the outer region, leading to an inward movement of the outer region particles.
This phenomenon, known as core-collapse \cite{Balberg:2002ue}, can impair tidal stripping.
Assuming a cross-section of $10~\rm cm^{2}/g$ and the absence of a host system, the onset timescale for the core collapse in the satellite halo is estimated to be approximately $16~\rm Gyr$ \cite{Balberg:2002ue, Koda:2011yb}.
However, tidal effects may halt or delay core collapse \cite{Nishikawa:2019lsc, Zeng:2021ldo}.
We examine the velocity dispersion profiles of the satellite halo and confirm that the core collapse does not occur in all our simulations.
Thus, the strengthening of DM self-interaction simply results in more particles from the inner region moving outward, leading to enhanced tidal stripping and effectively offsetting the impact of increasing orbital circularity.
The distribution of circularity $\eta$ and the ratio of pericenter distance to the viral radius have a peak around 0.6 and 0.37, respectively \cite{Jiang:2014zfa, vandenBosch:2017ynq}.
The pericenter distance of the $\eta=0.5$ orbit corresponds to the 20 percentile of the distribution.
Therefore, by adjusting the magnitude of the self-interaction cross-section, the issue of selecting extreme orbits can be systematically mitigated in the SIDM scenario.

As shown in Fig.~\ref{orbit}, the orbital period lengthens with increasing orbital circularity.
Nonetheless, in all of our simulations, DF2 experiences at least six pericenter encounters.
Increasing the orbital parameter $x_{\rm c}$ to 1.0 or further increasing circularity would result in only five pericenter passages within the simulation period.
After several attempts, we find that it is unlikely to generate a DMDG similar to DF2 from the initial condition we select, with only five pericenter passages, using a reasonable DM self-interaction cross-section.
Therefore, given the upper limit of the cross-section, the range of orbit selection remains restricted, and the mitigating capability of SIDM is limited.

\subsection{GC evolution in CDM halo}

The initial $r_{\rm GC}$ discussed in Section \ref{sub:sat} directly influences the distribution of GCs.
A GC population with a larger $r_{\rm GC}$ experiences stronger tidal force throughout the entire process.
Therefore, despite the weaker tidal effects in the case of CDM, we expect that a GC population with a larger $r_{\rm GC}$ can achieve a similar spatial distribution as the SIDM halo.

Fig.~\ref{cdmgc} shows the evolution of GCs in two CDM simulations with different $r_{\rm GC}$.
The spatial distribution of GCs in these two simulations exhibits opposite evolutionary trends.
In the simulation with $r_{\rm GC}=2.8~\rm kpc$, GCs are more strongly influenced by tidal effects due to their longer initial distance to the center of the satellite system.
The tidal effects have a greater impact on GCs than the dynamical friction, resulting in an increasingly diffuse GC population during the evolution.
While, in the simulation with $r_{\rm GC}=1.8~\rm kpc$, the dynamical friction is more pronounced and dominates the spatial distribution of GCs in the later stages of evolution.
Hence the half-number radius of GCs demonstrates a decreasing trend.
It can be seen that a diffuse GC population similar to that in DF2 can also be reproduced using an initial condition with a larger $r_{\rm GC}$ in the CDM scenario.
However, only 4 GCs remain within $10~\rm kpc$ in this simulation when the velocity dispersion decreases to a level consistent with the observational data.
GCs located at larger $r_{\rm GC}$ are more easily stripped away.
The lower efficiency in reducing velocity dispersion requires a longer time to reproduce GCs in DF2 compared to the SIDM case.
All these cause a low number of remaining GCs.
To ensure that the simulated satellite system ultimately contains 10 GCs, the initial number of GCs should be 25, far exceeding the value derived from the empirical relation between the number of GCs and halo virial mass \cite{Burkert:2019ucv}.
Therefore, although GCs in DF2 can also be replicated in the CDM case, extreme initial conditions are required.


\section{\label{sec:conclusion}Conclusion}

Utilizing controlled N-body simulations, we have verified that tidal evolution within an accreting host halo can transform a typical dwarf galaxy into a DMDG similar to DF2 in the SIDM scenario.
Our benchmark simulation has successfully replicated some crucial features of DF2, including its mass profile, the dispersion of stars, and the distribution of GCs.
We also performed an equivalent simulation in the standard CDM scenario for comparison.
The results indicate that the tidal stripping within a typical CDM halo is insufficient to reproduce the formation of DF2, suggesting that DF2 favors an explanation in terms of SIDM.
Likewise, the distribution of GCs is more likely to be realized in the case of SIDM.
Furthermore, we performed three additional simulations with varying combinations of DM self-interaction cross-section and orbital parameters, all of which generate a DMDG resembling DF2.
The enhanced tidal stripping due to DM self-interaction enables us to select an orbit with a higher pericenter distance.

\acknowledgments
We thank Daneng Yang and Yu-Ming Yang for helpful discussions. This work is supported by the National Natural Science Foundation of China under grant No. 12175248.


\nocite{*}

\bibliography{apssamp}

\end{document}